\documentclass[english]{article}
\usepackage[T1]{fontenc}
\usepackage[latin9]{inputenc}
\usepackage{geometry}
\usepackage{xcolor}
\usepackage{pdfcolmk}
\usepackage{babel}
\usepackage{amsmath}
\usepackage{amssymb}
\usepackage{graphicx}
\PassOptionsToPackage{normalem}{ulem}
\usepackage{ulem}
\usepackage[unicode=true,pdfusetitle,
 bookmarks=true,bookmarksnumbered=false,bookmarksopen=false,
 breaklinks=false,pdfborder={0 0 0},pdfborderstyle={},backref=false,colorlinks=false]
 {hyperref}

\makeatletter

\providecolor{lyxadded}{rgb}{0,0,1}
\providecolor{lyxdeleted}{rgb}{1,0,0}

\DeclareRobustCommand{\lyxsout}[1]{\ifx\\#1\else\sout{#1}\fi}

\newcommand{\lyxaddress}[1]{
	\par {\raggedright #1
	\vspace{1.4em}
	\noindent\par}
}

\usepackage{xcolor}
\hypersetup{
    colorlinks,
    linkcolor={red!50!black},
    citecolor={blue!50!black},
    urlcolor={blue!50!black}
}

\makeatother

\begin{document}
\title{From perturbative to non-perturbative in the \emph{O}(4) sigma model}
\author{Michael C. Abbott, Zoltán Bajnok, János Balog, Árpád Heged\H{u}s}
\maketitle

\lyxaddress{\begin{center}
~\\
Wigner Research Centre for Physics\\
Konkoly-Thege Miklós u. 29-33, 1121 Budapest, Hungary\\
\par\end{center}}
\begin{abstract}
We study the resurgent trans-series for the free energy of the two-dimensional
$O(4)$ sigma model in a magnetic field. Exploiting integrability,
we obtain very high-order perturbative data, from which we can explore
non-perturbative sectors. We are able to determine exactly the leading
real-valued exponentially small terms, which we check against the
direct numerical solution of the exact integral equation, and find
complete agreement.
\end{abstract}

\section{Introduction }

Perturbation theory in physics very seldom leads to a convergent Taylor
series in the coupling, but the divergent tail of this series contains
information about non-perturbative effects \cite{Bender:1969si,Brezin:1976wa}.
For many systems, it is now possible to map out a web of relations
between the original series expanding about the vacuum, and expansions
around other saddle points of the path integral, and the set of tools
for doing so is known as resurgence theory \footnote{For recent reviews see \cite{Marino:2012zq,Dorigoni:2014hea,Dunne:2015eaa,Aniceto:2018bis}.}.
Quantum field theories certainly have badly behaved perturbation theory
\cite{Dyson:1952tj,Hurst:1952zh,Lipatov:1976ny}, and contain non-perturbative
objects such as instantons and renormalons \cite{tHooft:1977xjm,Beneke:1998ui}.
But it is usually difficult to calculate enough terms to see the patterns
connecting them in much detail.

In the standard model of particle physics, perturbation theory works
extremely well for electroweak effects \cite{Aoyama:2012wk}. But
it is less useful for the strong force, where non-perturbative effects
such as quark confinement are unavoidable, and there is interest in
using these to make better predictions from perturbation theory \cite{Bauer:2011ws,Caprini:2020lff}.
The $O(N)$ sigma-model has often been used as a toy model for QCD,
exhibiting asymptotic freedom and a dynamical mass gap \cite{Hasenfratz:1990zz}.
We study in particular the $O(4)$ model, whose free energy (in terms
of the running coupling $\alpha$) has the following leading power-series
and exponential contributions, in a sense made preciese below:
\begin{align}
f & =1+\frac{\alpha}{2}+\frac{\alpha^{2}}{4}+\frac{10-3\zeta_{3}}{32}\alpha^{3}+\chi_{5}\alpha^{4}+\ldots\nonumber \\
 & \qquad+{\rm e}^{-8/\alpha}\,\big(d_{1}+d_{2}\alpha+d_{3}\alpha^{2}+\ldots\big)+\ldots.\label{eq:advert}
\end{align}
Only the first three perturbative terms $\chi_{n}$ have been calculated
by standard methods \cite{Bajnok:2008it}, but we can do much better
by exploiting the integrability of the $O(N)$ model \cite{Zamolodchikov:1977nu}.
In particular, a method for using its thermodynamic Bethe ansatz (TBA)
description \cite{Hasenfratz:1990zz} to calculate very high-order
perturbative coefficients $\chi_{n}$ was invented by Volin \cite{Volin:2009wr,Volin:2010cq}.
His 26 terms were sufficient to see the structure of the Borel plane,
where the leading singularities give rise to imaginary ambiguities
of order ${\rm e}^{-2/\alpha}$. We extend this work to calculate
2\,000 terms for the $O(4)$ case, for an energy $\epsilon$ and
density $\rho$ separately, with $f\propto\epsilon/\rho^{2}$. From
these we can map out the algebra of alien derivatives connecting different
non-perturbative sectors \cite{Dorigoni:2014hea}. These relations
ultimately allow us to recover the real ${\rm e}^{-8/\alpha}$ part
of the free energy \eqref{eq:advert}, and coefficients $d_{n}$,
via median resummation \cite{Marino:2008ya,Aniceto:2013fka}. We can
confirm this by comparing to a numerical solution of the TBA, which
includes all exponential corrections.\footnote{These and related calculations are described at greater length in
another article \cite{longWIP}.}

\section{Perturbation theory and TBA}

The $O(4)$ sigma model is a relativistic quantum field theory in
two dimensions, with four scalar fields $\Phi_{i}(x,t)$ restricted
to the unit sphere: $\sum_{i=1}^{4}\Phi_{i}^{2}=1$. When a magnetic
field $h$ is coupled to the conserved charge $Q_{12}$, the Lagrangian
reads \cite{Hasenfratz:1990zz}
\begin{align}
\mathcal{L} & =\smash{\frac{1}{2\lambda^{2}}\Big\{}\partial_{\mu}\Phi_{i}\partial^{\mu}\Phi_{i}+2ih(\Phi_{1}\partial_{0}\Phi_{2}-\Phi_{2}\partial_{0}\Phi_{1})\nonumber \\
 & \qquad\qquad+h^{2}(\Phi_{3}^{2}+\Phi_{4}^{2}-1)\Big\}.
\end{align}
One of the scalar fields may be eliminated, say $\Phi_{1}^{2}=1-\lambda^{2}(\varphi_{2}^{2}+\varphi_{3}^{2}+\varphi_{4}^{2})$
with $\lambda\varphi_{i}=\Phi_{i}$, and then the free energy density
$\mathcal{F}$ is given by the following path integral:
\begin{equation}
e^{-V\mathcal{F}(h)}=\int\mathcal{D}^{3}[\varphi]\:{\rm e}^{-\int d^{2}x\,\mathcal{L}(x)}.
\end{equation}
The density $\rho$ and the ground-state energy density $\epsilon(\rho)$
are related to $\mathcal{F}(h)$ by a Legendre transformation: 
\begin{equation}
\rho=-\partial\mathcal{F}/\partial h,\qquad\epsilon(\rho)=\mathcal{F}(h)-\mathcal{F}(0)+\rho h.
\end{equation}
Instead of standard perturbation theory in the bare coupling $\lambda$,
the expansion can be improved using the renormalization group, and
the free energy is eventually expressed in terms of the running coupling
$\alpha$, defined via
\begin{equation}
2/\alpha+1-\log\alpha=\log(\rho^{2}32\pi/m^{2}).\label{eq:defn-alpha}
\end{equation}
Direct perturbative results are available only for the first three
terms \cite{Bajnok:2008it}, and technically it is very difficult
to proceed to higher orders.

In the integrable description, the infrared degrees of freedom can
be used to calculate the ground state energy. A large enough magnetic
field forces these particles to condense into an interval $-B<\theta<B$
of rapidity, whose length depends on $h$. The thermodynamic limit
of the Bethe ansatz equation then leads to a linear integral equation
for the density of these particles, $\chi(\theta)$: 
\begin{equation}
\chi(\theta)-\int_{-B}^{B}\frac{d\theta'}{2\pi}K(\theta-\theta')\,\chi(\theta')=m\cosh\theta.\label{TBA}
\end{equation}
Here $K$ is the logarithmic derivative of the S-matrix
\begin{align}
2\pi K(\theta) & =-2\pi i\partial_{\theta}\log S(\theta)\\
 & =2\big\{\Psi(1-i\theta/2\pi)-\Psi(1/2-i\theta/2\pi)+\text{c.c.}\big\}\nonumber 
\end{align}
where $\Psi(\theta)=\partial_{\theta}\log\Gamma(\theta$) is the digamma
function. The density and energy are then
\begin{equation}
\rho=\int_{-B}^{B}\frac{d\theta}{2\pi}\chi(\theta),\qquad\epsilon=m\int_{-B}^{B}\frac{d\theta}{2\pi}\cosh\theta\,\chi(\theta).\label{rhoeps}
\end{equation}
The parameter $B$ can be related to the magnetic field by $h=\partial_{\rho}\epsilon(\rho)$,
which follows from minimizing the free energy over $\rho$. The large-$B$
expansion can thus be translated into a large-$h$ expansion, which
then can be compared to the original perturbative expansion. Such
a comparison was used to relate the dynamically generated $\Lambda_{\overline{MS}}$
scale to the masses of the particles \cite{Hasenfratz:1990zz}.

Volin's method to expand the TBA equation systematically works by
solving the TBA both in the bulk $\theta\sim0$ and near the edge
$\theta\sim B$, and then matching these two expansions, order by
order \cite{Volin:2009wr,Volin:2010cq}. Solving the recursion leads
to a large-$B$ expansion of both the ground-state energy
\begin{equation}
\epsilon=\hat{\epsilon}\:m^{2}{\rm e}^{2B}/16,\qquad\hat{\epsilon}=1+{\textstyle \sum_{k=1}^{\infty}}\:\xi_{k}\big/B^{k}
\end{equation}
and the density 
\begin{equation}
\rho=\hat{\rho}\:m\,{\rm e}^{B}\sqrt{B/8\pi},\qquad\hat{\rho}=1+{\textstyle \sum_{n=1}^{\infty}}\:u_{n}\big/B^{n}
\end{equation}
where we define $\hat{\epsilon}$ and $\hat{\rho}$ to standardise
on expansions starting with $1$.

He worked with generic $O(N)$ models, and was able to find the first
$26$ coefficients. These results were recently extended to $44$
coefficients in \cite{Marino:2019eym}, and to some non-relativistic
theories in \cite{Marino:2019fuy,Marino:2020dgc,Marino:2019wra}.
We decided to focus on the $O(4)$ model only, where we were able
to solve the recursive equations in closed form. This allowed us to
calculate $\sim50$ coefficients analytically, and 2\,000 coefficients
numerically with very high precision of 12\,000 decimal digits.

\section{Resurgence in $B$}

To explore the non-perturbative sectors and to reveal the resurgence
structure, we start with the density $\hat{\rho}$, whose first two
coefficients are
\begin{equation}
u_{1}=-\frac{3}{8}+\frac{\ell}{2},\qquad u_{2}=-\frac{15}{128}+\frac{3\ell}{16}-\frac{\ell^{2}}{8}
\end{equation}
where $\ell=\ln2$. From the first few coefficients, calculated analytically,
we observe that $u_{n}$ is a polynomial up to $\ell^{n}$, and may
contain zeta-functions, of odd order no higher than $n$. At large
$n$, we see that $u_{n}$ grows factorially, such that the following
$c_{n}$ approaches a constant:
\begin{equation}
c_{n}=2^{n+1}u_{n+1}/n!\,.\label{eq:cn}
\end{equation}
We have seen this with very high precision numerically \cite{longWIP}.
Technically we introduced $c_{n}$ by \eqref{eq:cn} and analysed
their large $n$ behavior via high order Richardson transform \cite{Aniceto:2018bis}.
The tail of the $100$th order Richardson transform was constant to
$150$ digits precision. The coefficients $\xi_{k}$ from the energy
$\hat{\epsilon}$ behave analogously.

\begin{figure}
\begin{centering}
\centering \includegraphics[width=10cm]{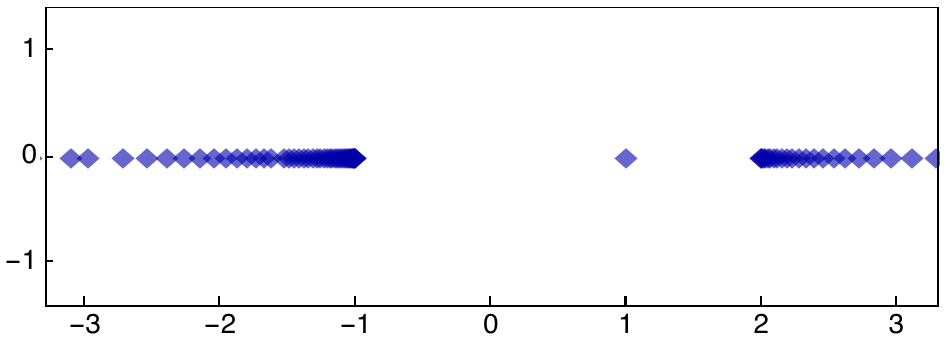}
\par\end{centering}
\caption{Positions of the poles of a 100th-order Padé approximant of the Borel
transform of $\hat{\epsilon}$, in the complex $t$ plane. These accumulate
along cuts $t\protect\leq-1$ and $t\protect\geq2$, plus an isolated
pole at $t=1$. \label{fig:Pade-poles-plot}}
\end{figure}

To see the analytic structure on the Borel plane (i.e. of the function
$\sum_{n}c_{n}t^{n}$) we plot the poles of the Padé approximant corresponding
to $\hat{\epsilon}$ in Fig. \ref{fig:Pade-poles-plot}. It shows
a cut starting at $t=-1$, a pole at $1$, and another cut starting
at $2$. The analytic structure of the Borel transform of $\hat{\rho}$
is similar, except without the pole at $t=1$. These agree with the
findings of \cite{Volin:2009wr,Marino:2019eym} for the free energy,
who established the factorial growth, and determined the location
of the cuts. They attributed this behaviour to UV and IR renormalons.

In order to see if our Borel transformed functions are simple resurgent
functions with logarithmic cuts we applied a two-step procedure as
in \cite{Aniceto:2018uik}. We first changed the asymptotics to ensure
a square root branch cut, then we used a conformal type mapping to
transform it into a pole singularity. The changes in the analytical
structures were followed by high order Pade approximants. These numericalresults
convinced us that we are dealing with simple resurgent functions and
all cuts are logarithmic, which we assume from now on.

The notion of an alien derivative for simple resurgent functions is
a concise and elegant way to characterize the logarithmic cut (and
pole) structure of the Borel transform. It is related to the logarithm
of the Stokes authormorphism, see later for our convention. We refer
to \cite{Dorigoni:2014hea} for the definition, and here merely summarize
the connection to asymptotic coefficients. Consider the formal asymptotic
expansion
\begin{equation}
\Psi(z)=1+{\textstyle \sum_{n=1}^{\infty}}s_{n}/z^{n},\qquad z=2B
\end{equation}
whose Borel transform is $B(t)=\sum_{n=0}^{\infty}c_{n}t^{n}$ with
$c_{n}=s_{n+1}/n!$ which behaves asymptotically as 
\begin{equation}
\begin{split}c_{n} & =\Big(p^{+}+\frac{p_{0}^{+}}{n}+\frac{p_{1}^{+}}{n(n-1)}+\dots\Big)\\
 & \quad+(-1)^{n}\Big(p^{-}+\frac{p_{0}^{-}}{n}+\frac{p_{1}^{-}}{n(n-1)}+\dots\Big)
\end{split}
\end{equation}
Then the alien derivatives at $t=\pm1$ are given by 
\begin{equation}
\Delta_{\pm1}\Psi(z)=\mp i2\pi\Big\{ p^{\pm}\pm\sum_{m=0}^{\infty}\frac{(\pm1)^{m}p_{m}^{\pm}}{z^{m+1}}\Big\}.\label{eq:Delta1}
\end{equation}

Treating $\hat{\rho}$ first, using a version of the Richardson transform
we could see with about 150 digits precision that all the $p^{+}$
coefficients vanish \cite{longWIP}. Technically we defined
the even $c_{2n}+c_{2n-1}$ and the odd $c_{2n}-c_{2n-1}$ combinations
and used a high order Richardson transform to read off $p^{+}$ and
$p^{-}$ with $150$ digits precision, respectively. We then subtracted
$p^{+}+(-1)^{n}p^{-}$ from $c_{n}$ and multiplied the result with
$n$ and repeated the analysis. Using these high precision numerical
values we found analytic expressions for the first $8$ coefficients
$p_{n}^{-}$, with similar structure to the original $u_{n}$ and
$\xi_{k}$ coefficients\footnote{This can be done e.g. by using the \texttt{FindIntegerNullVector }function
in Mathematica or by the EZ-Face website, CECM, Simon Fraser University:
\href{http://wayback.cecm.sfu.ca/projects/EZFace/}{http://wayback.cecm.sfu.ca/projects/EZFace/}}. We then determined the next 42 terms with high, but decreasing numerical
precision. After repeating the same analysis for $\hat{\epsilon}$,
and investigating the obtained coefficients of the alien derivatives
$\Delta_{\pm1}$we observed that they can be written in terms of the
original functions:
\begin{equation}
\begin{split}\Delta_{1}\hat{\rho} & =0,\\
\Delta_{1}\hat{\epsilon} & =-4i,
\end{split}
\qquad\qquad\begin{split}\Delta_{-1}\hat{\rho} & =i\hat{\epsilon}\hat{\rho},\\
\Delta_{-1}\hat{\epsilon} & =i\hat{\epsilon}^{2}.
\end{split}
\label{Deltapm1}
\end{equation}
Although we obtained this result for the first 40 coefficients we
believe it is true in general, which is a beautiful manifestation
of resurgence, and allows us to calculate the result of all combinations
of $\Delta_{\pm1}$ in terms of $\hat{\rho}$ and $\hat{\epsilon}.$

To study higher alien derivatives, we begin by observing that the
Borel transform of $1/\hat{\epsilon}$ has only a pole singularity
at $t=-1$, whose residue is exactly known. After removing this pole,
no singularity remains between $-2$ and $1$, thus we continue analytically
this function using Pade approximation and re-expand it around $t=-1/2$.
By this trick the large $n$ asymptotics of the new coefficients will
carry information not only about the singularity at $1$, but also
at $-2$ (being $3/2$ and $-3/2$ in the new variable). We then perform
a (rescaled by $3/2$) asymptotic analysis of these coefficients.
We found that the analogous $p^{-}$ coefficients vanished for 60
digits, implying $\Delta_{-2}(1/\hat{\epsilon})=0,$ i.e. there is
no cut there, and since alien derivatives obey the Leibniz rule, this
means that $\Delta_{-2}\hat{\epsilon}=0$.

Next define $G=(\hat{\epsilon}+\hat{\epsilon}^{\prime})/\hat{\rho}^{2}$,
where prime denotes $d/dz$. It is easy to see that $\Delta_{\pm1}G=0,$
hence its expansion around $t=0$ has radius of convergence $2$.
Applying the (rescaled by 2) asymptotic analysis we found numerically
that $\Delta_{-2}G=0$, which implies $\Delta_{-2}\hat{\rho}=0.$
Thus the point $t=-2$ does not seem to be singular for our model.
Using similar analysis we calculated 
\begin{equation}
\begin{split}\Delta_{2}\hat{\rho} & =iR/2,\qquad R=1+{\textstyle \sum_{n=1}^{\infty}}r_{n}/z^{n}.\end{split}
\end{equation}
Here we can fix the first 50 coefficients $r_{n}$ numerically, with
gradually decreasing precision. For the first five of these, we found
analytic expressions in terms of $\ell=\ln2$ and zeta-functions using
EZ-Face. The first three are:
\begin{equation}
\begin{split}r_{1} & =1/2+\ell,\quad\qquad r_{2}=-\ell/2-\ell^{2}/2,\\
r_{3} & =\frac{21}{64}+\frac{3}{4}\ell^{2}+\frac{\ell^{3}}{2}+\frac{3}{8}\zeta_{3}.
\end{split}
\end{equation}
Similar analyses also gave 
\begin{equation}
\begin{split}\Delta_{2}\hat{\epsilon} & =2iE,\qquad E=1+{\textstyle \sum_{n=1}^{\infty}}e_{n}/z^{n}\end{split}
\end{equation}
and again we fixed five coefficients exactly, including
\begin{equation}
\begin{split}e_{1} & =1/4,\quad\qquad e_{2}=5/32-\ell/2,\\
e_{3} & =\frac{57}{128}-\frac{5}{8}\ell+\ell^{2}.
\end{split}
\end{equation}
There appears to be no simple relation between these coefficients
$r_{n}$, $e_{n}$ and those of the original functions, $\hat{\rho}$
and $\hat{\epsilon}$. Thus unlike the resurgence at $\pm1$, which
is trivial in the sense that the functions resurge to themselves,
the resurgence at $t=2$ is non-trivial. In order to refer to this
fact we call $R$ and $E$ as second generation functions, while $\hat{\rho}$
and $\hat{\epsilon}$ as first generation ones. The naming also refers
to the fact that $\hat{\rho}$ and $\hat{\epsilon}$ can be obtained
directly from the expansion of the TBA equation \eqref{TBA}, while
this is not so clear for $R$ and $E$. The trivial resurgence at
$t=\pm1$ is manifested also for $E$ and $R$. Investigating $\eta=E/\hat{\rho}^{2}$,
we find numerically that $\Delta_{\pm1}\eta=0$, implying that 
\begin{equation}
\Delta_{1}E=0,\qquad\Delta_{-1}E=2i\hat{\epsilon}E.
\end{equation}
Similar analysis gives 
\begin{equation}
\Delta_{1}R=0,\qquad\Delta_{-1}R=i(4\hat{\rho}E+\hat{\epsilon}R).
\end{equation}
There is also some evidence (a few digits) of the vanishing of $\Delta_{-2}\eta$,
leading to 
\begin{equation}
\Delta_{-2}R=\Delta_{-2}E=0.
\end{equation}
which again demonstrates that the $t=-2$ point is not singular.

\begin{figure}
\begin{centering}
\centering \includegraphics[width=8cm]{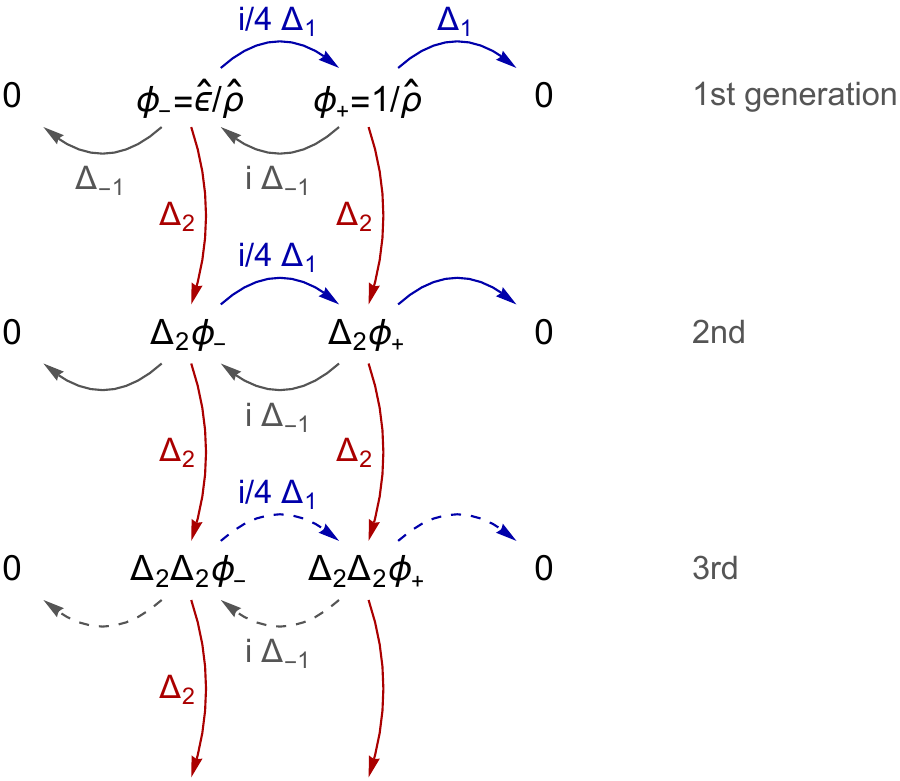}
\par\end{centering}
\caption{Resurgence pattern, in which alien derivative $\Delta_{1}$ maps to
the right, $\Delta_{-1}$ to the left, and $\Delta_{2}$ downwards,
while $\Delta_{-2}$ annihilates all functions. We have no information
about higher alien derivatives. The second generation contains $E,R$
and the third generation $\tilde{E},\tilde{R}$. Solid lines are links
we have demonstrated in \eqref{Deltapm1} to \eqref{eq:Delta2E};
dashed lines are conjectures, which allow us to fix $\Delta_{\pm1}\tilde{E}$
and $\Delta_{\pm1}\tilde{R}$. \label{fig:Generations}}
\end{figure}

Analysing the $t=2$ singularity we again observe a non-trivial resurgence with
new functions, which we call third generations 
\begin{equation}
\Delta_{2}E=-i/2\:\tilde{E},\qquad\Delta_{2}R=-i/2\:\tilde{R}\label{eq:Delta2E}
\end{equation}
where the leading expansions are
\begin{align}
\tilde{R} & =1+(\tfrac{1}{4}+\ell)/z+(\tfrac{5}{8}-\ell-2\ell^{2})/4z^{2}+{\rm \mathcal{O}}(1/z^{3})\nonumber \\
\tilde{E} & =1+3/8z^{2}+\mathcal{O}(1/z^{3}).
\end{align}
At this point we have run out of precision and we were not able to calculate alien derivatives $\Delta_{\pm3}$
or higher, but based on the available data we can conjecture
a pattern, shown in Fig. \ref{fig:Generations}. Defining
$\phi_{-}=\hat{\epsilon}/\hat{\rho}$ and $\phi_{+}=1/\hat{\rho}$
as representatives of the first generation, we observe that they are
exchanged by the action of $i/4\Delta_{1}$ and $i\Delta_{-1}$. Mapping
them to the second generation as $\Delta_{2}\phi_{\pm}$, we notice
that these functions are again exchanged by $\Delta_{\pm1}$, with
precisely the same coefficients. If this pattern persists, then it
fixes the third generation's $\Delta_{\pm1}\tilde{R}$ and $\Delta_{\pm1}\tilde{E}$.

\section{Resurgence in $\alpha$}

Our final goal is to recover the free energy $f=2\epsilon/(\alpha\pi\rho^{2})$
as calculated in traditional perturbation theory, \eqref{eq:advert}.
While expansions in large $z=2B$ are natural for the TBA calculation,
in perturbation theory the natural variable is the running coupling
$\alpha$. Thus we now switch to expanding in large $x=2/\alpha$,
related to $z$ by \cite{Volin:2009wr}
\begin{equation}
x{\rm e}^{x}=z{\rm e}^{z}4\hat{\rho}^{2}/{\rm e}.\label{plus}
\end{equation}
Since $\Delta_{\omega}$ is the alien derivative with respect to $1/z$
above, we write $D_{\omega}$ for alien derivatives of functions expanded
in $1/x$. We can translate between them using a formula for function
composition \cite{composit}: 
\begin{equation}
D_{\omega}\gamma(z(x))={\rm e}^{-\omega(z(x)-x)}(\Delta_{\omega}\gamma)\big(z(x)\big)+\gamma^{\prime}\big(z(x)\big)(D_{\omega}z)(x).
\end{equation}
Applying this to $\gamma=\hat{\rho}$ and combining it with the alien
derivative of (\ref{plus}) gives 
\begin{equation}
D_{\omega}\gamma=\Big(\frac{4z\hat{\rho}^{2}}{x{\rm e}}\Big)^{\omega}\Big\{\Delta_{\omega}\gamma-\frac{2x\dot{\gamma}}{(1+x)\hat{\rho}}\Delta_{\omega}\hat{\rho}\Big\}\label{starstarstar}
\end{equation}
where the dot indicates $d/dx$. We are interested in the alien derivatives
of $f=x\hat{\epsilon}/(z\hat{\rho}^{2})$, and we obtained
\begin{equation}
D_{\omega}f=\Big(\frac{4z\hat{\rho}^{2}}{x{\rm e}}\Big)^{\omega}\Big\{\frac{f}{\hat{\epsilon}}\Delta_{\omega}\hat{\epsilon}-\frac{2x(f+\dot{f})}{(1+x)\hat{\rho}}\Delta_{\omega}\hat{\rho}\Big\}.\label{starstar}
\end{equation}
Due to the special form of the relation between $x$ and $z$, \eqref{plus},
the singularities in the $x$ variable are at the same positions as
they are in $z$. After a long calculation we obtained that 
\begin{equation}
\begin{split}D_{1}f & =-16i/{\rm e},\\
D_{2}f & =\frac{16i}{{\rm e}^{2}}{\cal F},\qquad{\cal F}=\frac{2z\hat{\rho}^{2}}{x}E-\frac{z^{2}\hat{\rho}^{3}(f+\dot{f})}{x(1+x)}R.
\end{split}
\label{eq:D2eff}
\end{equation}
Using (\ref{starstarstar}), it follows that $D_{1}{\cal F}=D_{-2}{\cal F}=0$.
We can similarly calculate $D_{-1}f$ and $D_{-2}f=0$, and then $D_{-1}\mathcal{F}$.
In order to make comparision with the TBA result we will also need
\begin{equation}
D_{2}{\cal F}=\frac{16i}{{\rm e}^{2}}\Big(1-\frac{5}{4x}-\frac{1}{2x^{2}}+\dots.\Big).\label{eq:D2curlyF}
\end{equation}

\section{Median resummation}

Using these alien derivatives of $f=\sum_{n=1}\chi_{n}\alpha^{n}$,
we can now propose an ambiguity-free resummation of the perturbative
series. Clearly the two lateral Borel resummations 
\begin{equation}
S_{\pm}(f)=\chi_{1}+\alpha\chi_{2}+\smash{\int_{0}^{\infty\pm i0}}dt\:{\rm e}^{-tx}B(t)
\end{equation}
are different, due the singularities on the positive real line. They
are related by the Stokes automorphism $\mathfrak{S}$, which can
be written in terms of the alien derivatives as\footnote{Observe that we introduced an unconventional extra sign in the relation
between the alien derivative and the Stokes automorphism. This is
also consistent with the large order relations \eqref{eq:Delta1}.}
\begin{equation}
S_{+}(f)=S_{-}(\mathfrak{S}f),\qquad\mathfrak{S}=\exp\Big(-\sum_{n=1}^{\infty}{\rm e}^{-nx}D_{n}\Big).
\end{equation}
The median resummation arises from demanding that the lateral resummations
for the trans-series are the same, and it involves the square root
of the Stokes automorphism \cite{Marino:2008ya,Aniceto:2013fka,Dorigoni:2014hea}. 
It was formulated
originally for a one-parameter trans-series, but we assume that it
works also for our more complicated case.
In terms of the alien derivatives the proposed median
resummation reads as:
\begin{equation}
S_{\mathrm{med}}(f)=S_{+}(\mathfrak{S}^{-\frac{1}{2}}f)=S_{+}\big({\rm e}^{\frac{1}{2}\sum_{n=1}^{\infty}e^{-nx}D_{n}}f\big).
\end{equation}

\begin{figure}
\begin{centering}
\includegraphics[width=10cm]{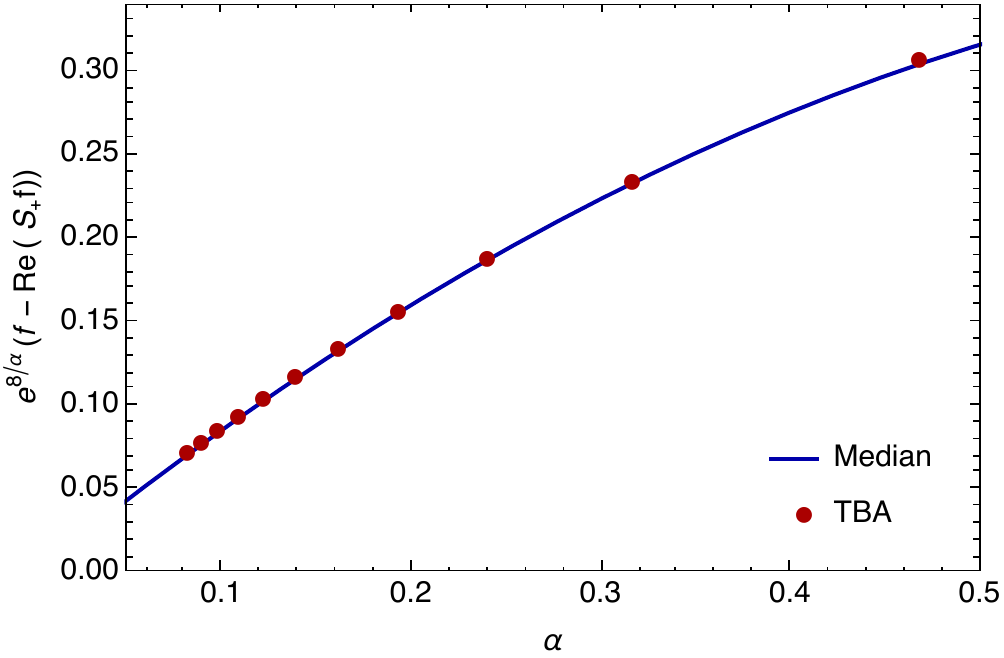}
\par\end{centering}
\caption{Comparison of the numerical solution of the TBA (performed at $B=2,3,\ldots12$,
to at least 40 digits) with the results from the median resummation,
truncated to terms shown in \eqref{final} and \eqref{eq:D2curlyF}.
We subtract from both the averaged lateral Borel resummation $\text{Re}(S_{+}(f))$,
and divide by ${\rm e}^{-8/\alpha}$. The numerical errors in both
data are negligibly small $(<10^{-20})$ compared to their deviations.
The points at $\alpha<0.4$ (i.e. $B\protect\geq3$) agree to 3 digits
and the deviation can be attributed to the neglected higher order terms in $D_{2}D_{2}f$
and start at $\alpha^{3}.$}

\label{fig: tba}
\end{figure}

Expanding this, notice that $D_{1}f=-16i/{\rm e}$ implies that all
higher terms $D_{k}D_{1}f$ vanish. We also found that $D_{1}D_{2}f=0$,
and so the leading terms are:

\begin{equation}
\begin{split}S_{{\rm med}}(f)= & S_{+}\Big(f+\frac{{\rm e}^{-x}}{2}D_{1}f+\frac{{\rm e}^{-2x}}{2}D_{2}f+\frac{{\rm e}^{-3x}}{2}D_{3}f+\frac{{\rm e}^{-4x}}{2}D_{4}f+\dots\\
 & +\frac{{\rm e}^{-4x}}{8}D_{2}D_{2}f+\frac{{\rm e}^{-4x}}{8}D_{1}D_{3}f+\dots\Big),
\end{split}
\label{eq:Smed-1}
\end{equation}
where the dots stand for contributions proportional to ${\rm e}^{-5x}$
and higher.

On the other hand, using the definition of the Stokes automorphism
we write 
\begin{equation}
S_{-}(f)=S_{+}({\mathfrak{S}}^{-1}f)=S_{+}\big({\rm e}^{\sum_{n=1}^{\infty}{\rm e}^{-nx}D_{n}}f\big)
\end{equation}
and expanding this up to O$({\rm e}^{-4x})$ we obtain 
\begin{equation}
\begin{split}S_{-}(f)= & S_{+}\Big(f+{\rm e}^{-x}D_{1}f+{\rm e}^{-2x}D_{2}f+{\rm e}^{-3x}D_{3}f+{\rm e}^{-4x}D_{4}f+\dots\\
 & +\frac{{\rm e}^{-4x}}{2}D_{2}D_{2}f+\frac{{\rm e}^{-4x}}{2}D_{1}D_{3}f+\dots\Big).
\end{split}
\end{equation}

We have not calculated $D_{3}f$ and $D_{4}f$, however, these unknown
quantities drop out from the formula if we express $S_{{\rm med}}(f)$
using the averaged lateral resummation 
\begin{equation}
{\rm Re}\big(S_{+}(f)\big)=\frac{1}{2}S_{+}(f)+\frac{1}{2}S_{-}(f).
\end{equation}
The median resummation formula simplifies to 
\begin{equation}
S_{{\rm med}}(f)={\rm Re}\big(S_{+}(f)\big)-\frac{{\rm e}^{-4x}}{8}D_{2}D_{2}f-\frac{{\rm e}^{-4x}}{8}D_{1}D_{3}f+\dots
\end{equation}
We have not (yet) calculated $D_{1}D_{3}$ either, but since we have
seen that for almost all of our functions the alien derivative at
$+1$ vanishes, at this point we make the bold assumption 
\begin{equation}
D_{1}D_{3}f=0. \label{eq:D1D3}
\end{equation}
With this assumption the final result reads 
\begin{equation}
S_{{\rm med}}(f)={\rm Re}\big(S_{+}(f)\big)-\frac{{\rm e}^{-4x}}{8}D_{2}D_{2}f+\dots\label{final}
\end{equation}
We have investigated this formula (and hence our assumption) numerically,
by comparing it to the result obtained from numerically solving the
TBA equations. To calculate the integral $S_{+}(f)$, we used the
conformal mapping method to obtain a precise enough analytical continuation
of $B(t)$, which we integrated numerically with high (at least 20
digits) precision \cite{longWIP}. We see convincingly good
agreement, shown in Fig. \ref{fig: tba}, which indicates that our assumption \eqref{eq:D1D3} might be correct. 
It would be very
nice to calculate also these missing terms in the future. Our numerical precision
is sufficient to clearly see some deviation between the exact TBA
result and that given by (\ref{final}) but this can be attributed
to the missing O$(\alpha^{3})$ (and higher) terms in \eqref{eq:D2curlyF}.

\section{Conclusion}

We investigated the free energy of the two dimensional $O(4)$ sigma
model in a magnetic field. Using the integrability of the model we
determined 2,000 perturbative coefficients with very high precision,
which enabled us to investigate the analytic structure of the density
and energy on the Borel plane. Using asymptotic analyses we revealed
a nice resurgence structure and determined via the median resummation
the leading exponentially small corrections to the free energy \eqref{eq:advert},
with $d_{0}=32/e^{4}$, $d_{1}=-20/e^{4}$ and $d_{2}=-4/e^{4}$,
which agreed with the numerical solution of the TBA equation. These
results fit into a trans-series of the form 
\begin{equation}
f=\sum_{m=0}^{\infty}e^{-2m/\alpha}\Big(\sum_{n=1}^{\infty}\chi_{n}^{(m)}\alpha^{n-1}\Big)
\end{equation}
where $\chi_{n}^{(0)}=\chi_{n}$ are the perturbative coefficients,
$\chi_{n}^{(1)}=-8i/{\rm e}^{2}\delta_{n,1}$ are related to $D_{1}f$,
and $\chi_{n}^{(2)}$ to $D_{2}f$. Similar formulations are possible
for each of $\rho$ and $\epsilon$. For more general observables
and complex couplings we expect a trans-series with exponents $e^{2m/\alpha}$
corresponding to the negative alien derivatives. These cannot fit
into the best-studied case of a one-parameter trans-series, because
of the more complicated pattern of resurgence relations we have discovered. It
would be interesting to formulate {\'E}calle's bridge equations for
this theory.

It would also be interesting to extend this work to other $O(N)$
models, or to similar theories \cite{Marino:2019eym,Kazakov:2019laa}.
The method of \cite{Volin:2009wr} for calculating $\chi_{n}^{(0)}$
works for all $O(N)$, but is more complicated. It may also be possible
to extract information about higher $\chi_{n}^{(m)}$ directly from
the TBA, instead of starting only from perturbative data.

Finally (and more importantly), it would be interesting to know what
semi-classical configurations (instantons, renormalons, bions \cite{Fujimori:2016ljw,Fujimori:2018kqp})
are responsible for the resurgence structure found here if there is
any.

\subsection*{Acknowledgements}

We thank Ines Aniceto and Daniel Nogradi for useful discussions. Our
work was supported by ELKH, with infrastructure provided by the Hungarian
Academy of Sciences. This work was supported in part by NKFIH grant
K134946. MCA was also supported by NKFIH grant FK128789.

\end{document}